\documentclass[10pt]{article}

\makeatletter
\def\@@M{{\mathrm I \mkern-2.5mu \nonscript\mkern-.5mu \mathrm I
\mkern-5.5mu\mathrm  M}}
\def\@@N{{\mathrm I \mkern-2.5mu \nonscript\mkern-.5mu \mathrm I
\mkern-5.5mu\mathrm  N}}
\def\@@Z{{
  \setbox0\hbox{\sf Z}
  \setbox1\hbox{$\mathrm\kern.05\wd0$
                \rlap{\vrule height.93\ht0 depth-.75\ht0 width.056\wd0 }%
                \kern-.13\wd0 \copy0 \kern-.6\wd0 \copy0 \kern-.1\wd0
                \llap{\vrule height.25\ht0 depth\z@ width.056\wd0}%
                \kern.05\wd0}
  \mathchoice{\copy1}{\copy1}{\mit Z\mkern-8mu Z}{\mit Z\mkern-7.5mu Z} }}
\def\@@insvline#1#2{{\setbox0\hbox{\m@th$#1\mathrm I$}
  \rlap{\m@th$#1 \mkern 5mu
  \vrule height.92\ht0 depth-.05\ht0 width.09\ht0 $}
  {\mathrm #2} }}
\def\@@Q{\mathpalette\@@insvline{Q}}
\def\@@R{{\mathrm I \mkern-2.5mu \nonscript\mkern-.5mu \mathrm R}}
\def\@@C{\mathpalette\@@insvline{C}}
\def\@@K{{\mathrm I \mkern-2.5mu \nonscript\mkern-.5mu\mathrm  K}}
\def\@@D{{\mathrm I \mkern-2.5mu \nonscript\mkern-.5mu\mathrm D}}

\def\callig#1{
  \ifcat#1a
    \ifnum`#1=\uccode`#1 {\cal #1}
    \else                #1
    \fi
  \else                  #1
  \fi}

\def\varbbbold#1{
  \ifcat#1a
    \ifnum`#1=\uccode`#1 \csname @@#1\endcsname
    \else                #1
    \fi
  \else                  #1
  \fi}

\def\eulerfraktur#1{
  \ifcat#1a {\frak #1}
  \else                  #1
  \fi}
 
\newif\if@@excloccurred
\newif\if@@questoccurred

\let\@@quest=? \catcode`?=\active
\let\@@excl=!  \catcode`!=\active \let\fact\@@excl

\def!{\protect\p@@doexcl}
\def?{\protect\p@@doquest}

\def\p@@doexcl{\let\@@xeq\relax\ifmmode\def\@@xeq{\futurelet\@@next\@@doexcl}
                               \else\@@excl\fi\@@xeq}
\def\p@@doquest{\let\@@xeq\relax\ifmmode\def\@@xeq{\futurelet\@@next\@@doquest}
                                \else\@@quest\fi\@@xeq}
\def\@@doexcl{\let\@@x@q\relax
     \if@@excloccurred \@@excloccurredfalse
                       \ifx\@@next\@sptoken\@@excl\@@excl
                       \else\let\@@x@q\doexclexcl\fi
\else\if@@questoccurred\@@questoccurredfalse
                       \ifx\@@next\@sptoken\@@quest\@@excl
                       \else\let\@@x@q\doquestexcl\fi
\else\ifx\@@next\@sptoken\@@excl
\else\ifx\@@next!\@@excloccurredtrue
\else\ifx\@@next?\@@excloccurredtrue
\else\let\@@x@q\doexcl\fi\fi\fi\fi\fi\@@x@q}

\def\@@doquest{\let\@@x@q\relax
     \if@@excloccurred \@@excloccurredfalse
                       \ifx\@@next\@sptoken\@@excl\@@quest
                       \else\let\@@x@q\doexclquest\fi
\else\if@@questoccurred\@@questoccurredfalse
                       \ifx\@@next\@sptoken\@@quest\@@quest
                       \else\let\@@x@q\doquestquest\fi
\else\ifx\@@next\@sptoken\@@quest
\else\ifx\@@next!\@@questoccurredtrue
\else\ifx\@@next?\@@questoccurredtrue
\else\let\@@x@q\doquest\fi\fi\fi\fi\fi\@@x@q}

\def\doexcl{\bold}        \def\doquest{\callig}
\def\doexclexcl{\varbbbold} \def\doexclquest{\boldsymbol}
\def\doquestexcl{\Bbb}    \def\doquestquest{\eulerfraktur}
\makeatother

\advance\hoffset by -7mm
\makeatletter
\@addtoreset{equation}{section}
\makeatother

\renewcommand{\title}[1]{\null\vspace{25mm}

\noindent{\Large{\bf #1}}\vspace{10mm}

\noindent }
\newcommand{\authors}[1]{\noindent{\large #1}\vspace{3mm}

}
\newcommand{\address}[1]{\noindent #1\vspace{5mm}

}
\renewcommand{\abstract}[1]{\vspace{19mm}

\noindent{\small{\em Abstract.} #1}\vspace{2mm}

}

\usepackage{amstex}
\usepackage{enumerate}
\usepackage{ctagsplt}

\newtheorem{proposition}{Proposition}
\newtheorem{theorem}{Theorem}
\newtheorem{lemma}{Lemma}

\begin{document}

\begin{flushright}
ZU-TH-12/96\\
gr-qc/9604054
\end{flushright}

\title{On Newton-Cartan Cosmology \footnotemark }

\footnotetext{Dedicated to the sixtieth birthdays of Klaus Hepp and Walter
 Hunziker (to appear in Helv.~Phys.~Acta)}

\authors{Christian R\"uede and Norbert  Straumann}

\address{Institute of Theoretical Physics, University of Z\"urich, 
Winterthurerstrasse 190, CH-8057 Z\"urich, Switzerland}

\abstract{After a brief summary of the Newton-Cartan theory in a form which
 emphasizes its close analogy to general relativity, we illustrate the theory
 with selective applications in cosmology. The geometrical formulation of this
 nonrelativistic theory of gravity, pioneered by Cartan and further  developed
 by various workers, leads to a conceptually sound basis of Newtonian
 cosmology. In our discussion of homogeneous models and cosmological
 perturbation theory, we stress the close relationship with their general
 relativistic treatments. Spatially compact flat models also fit into this
 framework.}

\section{Introduction}

We hope that Klaus and Walter will accept this modest note as a tribute to
 their outstanding role as teachers of theoretical physics. In their courses
 they present not only elegant techniques and formal developements, but always
 emphasize the importance of basic concepts. ``Rechnen kann jeder'', as
 Heitler used to say. In this spirit, we devote this article to a theme which
 is mainly of conceptual nature, and -- as we hope -- also of some pedagogical
 interest.

We shall try to make it apparent that Newtons theory of gravity is much closer
 to general relativity (GR) than commonly appreciated. This has often been
 stressed in private conversations and letters by our inspiring teacher and
 colleague Markus Fierz. Here an example from a letter (Nov. 22, 1993):

\begin{quote}
,,Es war  um 1953, als ich meinen Newton-Aufsatz schrieb, dass ich Pauli
 sagte, auch in der allgemeinen Relativit\"atstheorie seien Raum-Zeit 
 ,absolut`, wie bei Newton. Darauf antwortete Pauli zu  meinem Staunen: ,Sie
 verraten damit, das Grundprinzip der allgemeinen Relativit\"atstheorie: dass
 n\"amlich Raum-Zeit-Materie nicht unabh\"angig voneinander gedacht werden
 k\"onnen`. Wir konnten uns dann aber einigen, indem ich zugab, dass hier das
 ,sine mutua actione` Newtons nicht  gilt, obwohl im Ganzen  diese
 Wechselwirkung klein ist. Was ich im Sinne hatte ist dies: Leibniz
 erkl\"arte, der Raum sei nichts Wirkliches, sondern entspringe der Ordnung
 der Monaden (Kraft pr\"astabilisierter Harmonie). Newton erwiderte hierauf:
 der Raum sei mehr als eine blosse Ordnung. Denn der Abstand zweier Punkte
 habe einen Sinn ganz unabh\"angig davon, ob der Raum von etwas erf\"ullt oder
 leer sei. Newton legte also grosses Gewicht auf den metrischen Charakter des
 Raumes: dieser macht ihn zum Gegenstand der Physik, zu etwas Wirklichem
 \dots \nolinebreak``
\end{quote}
Every honest teacher of theoretical physics is confronted at a very early
 stage of a classical mechanics course with the following difficulty: After
 having introduced -- in the spirit of L.~Lange -- the operational definition
 of an inertial frame, the question arises how to proceed when gravitational
 fields are present. In the traditional presentation of Newtons theory one
 maintains the fiction of an integrable (flat) affine connection, and puts
 gravity on the side of the forces, described by vector fields. A much more
 satisfactory formulation was given by Cartan \cite{cartan1} and Friedrichs
 \cite{friedrichs}. This denies the separate existence of a flat affine
 connection of space-time and a vector field describing gravitation, but puts
 gravity on the side of a more general dynamical connection which represents
 both inertia and gravitation.

Historically, this important step of course was  made first  by Einstein when
 he created his general theory of relativity, but it is clearly independent of
 the relativization of time. Following Cartan and Friedrichs, numerous authors
 have elaborated on this idea. Here we mention only a selective list of
 contributions by Havas \cite{havas}, Trautman \cite{trautman}, Ehlers
 \cite{ehlers1, ehlers2, ehlers3} and K\"unzle \cite{kuenzle1, kuenzle2}.
\\\\
In the first part of the present paper we give a brief summary of the
 Newton-Cartan theory, following mainly the work of H.P.~K\"unzle, a former
 diploma student of Fierz at the ETH. K\"unzle's presentation, which uses the
 language of fibre bundles, appears to us as the most natural one, because it
 just replaces the role of the Lorentz group in GR by the Galilei group.
 Following this route, one arrives at a theory which is slightly more general
 than Newton's theory. The latter is only obtained after imposing a somewhat
 strange looking nonlinear condition for the Riemann tensor. The structural
 analogy of GR and the Newton-Cartan theory is, however, striking. In
 particular, the field equations look identical.

In later sections we shall illustrate this also in more concrete terms with
 some selective applications in cosmology (homogeneous cosmological models and
 cosmological perturbation theory). This is perhaps not only an academic
 exercise, because much of the activity in cosmology, especially in connection
 with large scale structure formation, relies on the Newtonian approximation.
 We take this as a motivation for putting Newtonian cosmology on a
 conceptually firm basis. This has to be regarded as an extension of classic
 works by Heckmann \cite{heckmann} and Heckmann and Sch\"ucking
 \cite{heckmannschuecking}. One advantage which results is the possibility to
 choose the spatial sections as flat tori and thus describe compact
 cosmologies.
\\\\
All this confirms in more technical terms the remarks by Markus Fierz, quoted
 earlier. Einstein was wrong when he believed that his theory of gravitation
 incorporated the principle of Mach which is entirely in the spirit of
 Leibniz. This became already quite clear with the famous solution of G\"odel,
 but some relativists, notably Einstein himself, maintained the belief that
 Mach's principle might have something to do with the finiteness of space
 \cite{einstein1}. That this is not the case was once and for all demonstrated
 by the ``finite rotating universe'' solution found by Ozv\`ath and
 Sch\"ucking \cite{ozvathschuecking}. Space-time has really an independent
 existence and we are in fact still much closer to Newton than to Leibniz.

A more detailed account of the material treated in  this paper can be found in
 the diploma work by one of us \cite{rueede}.

\section{Galilei spacetimes and their connections}

In what follows, $M$ will always denote the space-time manifold and $L(M)$ the
 principle bundle of linear frames with the structure group $GL(4,!!R)$. In GR
 space-time is endowed with a Lorentz metric $g$ which defines a bundle
 reduction of $L(M)$ to the orthonormal frame bundle $O(M)$ with the
 homogeneous Lorentz group as the structure group. Conversely, each reduction
 of the structure group $GL(4,!!R)$ to the homogeneous Lorentz group gives
 rise to a Lorentz metric, because any element $u \in L(M)$ over $x \in M$ can
 be regarded as a linear isomorphism of $!!R^{4}$ onto $T_{x}M$, which maps
 the standard basis $\lbrace e_{\mu}\rbrace$ of $!!R^{4}$ to the linear frame
 $u$.

In a ``nonrelativistic'' gravity theory $M$ has to be endowed with a
 {\bf Galilei metric}, which consists of a one-form $\tau$ and a symmetric
 semi-definit contravariant tensor field $h$ of rank 3, satisfying
 $h(\cdot,\tau) = 0$ ($h^{\mu \nu}\tau_{\nu} = 0$). The pair $(h,\tau )$
 defines again a bundle reduction of $L(M)$, this time with the homogeneous
 Galilei group as structure group. The reduced bundle consists of all frames
 $\lbrace e_{\nu }\rbrace$ in $L(M)$, satisfaying
\begin{equation}\label{1}
\tau(e_{0}) = 1, \qquad h(\theta^{\mu},\theta^{0}) = 0, \qquad
h(\theta^{i},\theta^{j}) = \delta^{ij} \qquad (i,j = 1,2,3),
\end{equation}
where $\lbrace \theta^{\mu}\rbrace$ denotes the dual frames,
 $\langle \theta^{\mu}, e_{\nu}\rangle = \delta^{\mu}_{\nu}$.
 Since $h(\theta^{\mu}, \tau ) = 0$, these equations imply
 $\tau = \theta^{0}$. 

Conversely, a reduction of the structure group $GL(4,!!R)$ to the homogeneous
 Galilei group gives rise to a Galilei metric $(h,\tau)$. This just reflects
 the fact that the homogeneous Galilei group (without time reflections) is the
 subgroup of $GL(4,!!R)$ which leaves the standard Galilei metric of $!!R^{4}$
 invariant. The latter is defined by equations (\ref{1}) for the standard
 basis $\lbrace e_{\mu}\rbrace$ of $!!R^{4}$ and its dual. This defines the
 {\bf flat} Galilei spacetime. With this notion it is also clear what is ment
 by {\bf locally flat} Galilei spacetimes. These can be characterized as
 follows.

\begin{proposition} \label{prop1}
A Galilei spacetime $(M,h,\tau)$ is locally flat iff the following two
 conditions are satisfied:
\begin{enumerate}[{\em (i)}]
\item $d\tau = 0$;
\item the induced Riemannian metrics on the integral manifolds defined by
 $\tau$ are locally flat.
\end{enumerate}
\end{proposition}
From now on we shall only consider bundle reductions to the identity
 components $G_{+}^{\uparrow}$ of the homogeneous Galilei group (orthochronous
 Galilei group), which we shall denote by $\mathcal{G}(M,G_{+}^{\uparrow})$.
 The corresponding frames are then space and time oriented. 

It is now clear, how to define a {\bf Galilei connection} on $(M, h, \tau)$.
 This is a connection in the corresponding principle bundle
 $\mathcal{G} (M,G_{+}^{\uparrow})$, which we describe by a connection form
 $\omega$, satisfying the usual conditions. There is a natural
 characterization of Galilei connections:

\begin{proposition} \label{prop2}
A linear connection $\Gamma$ on a Galilei manifold $(M,h,\tau)$ is a Galilei
 connection iff
\begin{equation} \label{2}
\nabla h=0, \qquad \qquad \nabla \tau = 0,
\end{equation}
where $\nabla$ denotes the covariant derivative with respect to $\Gamma$.
\end{proposition}
We consider only symmetric connections. For these the second equation in
 (\ref{2}) implies $d\tau =0$. Thus the distribution defined by the 1-form
 $\tau$ is integrable. The corresponding maximal integral manifolds are the
 spatial {\bf sections of constant time}. Vectors tangent to these sections
 are annihilated by $\tau$ and are called {\bf spacelike} (or horizontal).
 Tangent vectors which are not annihilated by $\tau$ are called
 {\bf timelike}. If $\tau (V)=1$ we say that $V$ is a
 {\bf timelike unit vector}.

In contrast to Lorentz manifolds there is {\em no unique} symmetric Galilei
 connection on a Galilei manifold. It is instructive to see this in the light
 of a famous theorem by Weyl \cite{weyl} and Cartan \cite{cartan2}. Since this
 is not so well-known (even among relativists) we state it here:

\begin{theorem}
{\em (Weyl, Cartan)}~~ For a closed subgroup $G$ of $GL(n,!!R)$, $n \geq 3$,
 the following two conditions are equivalent:
\begin{enumerate}[{\em (i)}]
\item $G$ consists of all elements of $GL(n,!!R)$ which preserve a certain
 non-degenerate quadratic form of any signature;
\item For every $n$-dimensional manifold $M$ and for every reduced subbundle
 $P$ of $L(M)$ with group $G$, there exists a unique torsion-free connection
 in $P$.
\end{enumerate}
\end{theorem}
It turns out that the set of symmetric Galilei connections is in 1:1
 correspondence with the set $\Lambda^{2}(M)$ of $2$-forms on $M$. The two
 equations (\ref{2}) imply that the difference of two connection forms
 (Christoffel symbols) is given by a tensor field of the following type
\begin{equation} \label{3}
S^{\mu}_{\alpha \beta} = 2\tau_{(\alpha} \kappa_{\beta)\lambda}
 h^{\lambda \mu},
\end{equation}
where $\kappa_{\alpha \beta}$ are the components of a $2$-form $\kappa$.
Special symmetric Galilei connections can be described as follows. Choose a
 timelike unit vector field $V$ and define the covariant metric $h^{\flat}$
 (relative to $V$) such that its components $h_{\mu \nu}$ satisfy
\begin{equation} \label{4}
h_{\mu \nu}V^{\nu} =0, \qquad \qquad  h_{\mu \lambda}h^{\lambda \nu} =
 \delta ^{\nu}_{\mu} - \tau_{\mu}V^{\nu},
\end{equation}
then
\begin{equation} \label{5}
^{V}\Gamma^{\mu}_{\alpha \beta} = h^{\mu \rho}(V^{\sigma} \!\!  _{,\rho}
 h_{\sigma (\alpha}\tau_{\beta )} + \frac{1}{2}
 h^{\lambda \sigma} \!\!  _{,\rho} h_{\lambda \alpha} h_{\sigma \beta}) -
    h_{\rho (\alpha} h^{\rho \mu} \!\!  _{,\beta} - \tau_{(\alpha}
 V^{\mu} \!\! _{,\beta )}
\end{equation}
defines a symmetric Galilei connection. This is actually the unique symmetric
 Galilei connection which satisfies also
\begin{equation} \label{6}
V^{\rho}V^{\alpha} \!\! _{;\rho} = 0, \qquad \qquad  h^{\rho \alpha} V^{\beta}
 \!\! _{;\rho}  -  h^{\rho \beta} V^{\alpha} \!\! _{;\rho} =0.
\end{equation}
With respect to (\ref{5}) the vector field $V$ is then geodesic and rotation
 free.

Note that relative to a Galilei frame with $V = e_{0}$, equations (\ref{4})
 reduce to $h_{\mu 0} = 0$ and $h_{ij} = \delta_{ij}$; thus
 $h^{\flat} = \delta _{ij} \theta^{i} \otimes \theta^{j}$. This is a
 Riemannian metric on the leaves of the foliation defined by $\tau$. Clearly,
 the restriction of $h^{\flat}$ on an integral manifold is independent of $V$,
 because this is just the inverse of the restriction of the metric $h$.

One can show that the integral manifolds (sections of constant time) are
 totally geodesic for any symmetric Galilei connection and that the induced
 connection on a leave coincides with the Levi-Civit\`a connection
 corresponding to $h^{\flat}$.

The Newton-Cartan theory of gravity involves special symmetric Galilei
 connections of the form 
\begin{equation} \label{7}
\Gamma^{\mu}_{\alpha \beta} =\  ^{V}\! \Gamma^{\mu}_{\alpha \beta} +
 S^{\mu}_{\alpha \beta},
\end{equation}
where $S^{\mu}_{\alpha \beta}$ is given by (\ref{3}) with $d\kappa = 0$. Such
 connections will be called {\bf Newtonian}.

We need also a characterization of locally flat Galilei spacetimes.

\begin{proposition} \label{prop3}
For a Galilei manifold $(M,h,\tau)$ with symmetric Galilei connection $\Gamma$
 the following statements are equivalent:
\begin{enumerate}[{\em (i)}]
\item the Galilei manifold is locally flat;
\item $R^{\mu \nu} := h^{\mu \rho} h^{\nu \sigma} R_{\rho \sigma} = 0$;
\item $R_{\mu \nu} = \alpha_{(\mu} \tau_{\nu )} \qquad $ for some 1-form
 $\alpha$.
\end{enumerate}
\end{proposition}
Recalling that any Newtonian connection can be expressed in terms of the
 Galilei metric $(h,\tau)$, a timelike unit vector field $V$ and a closed
 $2$-form $\kappa$, the question arises, when -- for a given Galilei
 metric -- a change of $\kappa$ and $V$ does not affect the Newtonian
 connection. One can show that this {\bf Newtonian gauge group} is given by 
\begin{equation} \label{8}
\begin{aligned}
V^{\mu} & \longmapsto V^{\mu} + h^{\mu \nu} w_{\nu}, \\
A_{\mu} & \longmapsto A_{\mu} + f_{,\mu} + w_{\mu} - (V^{\nu} w_{\nu} +
 \frac{1}{2} h^{\nu \lambda} w_{\nu} w_{\lambda}) \tau_{\mu},
\end{aligned}
\end{equation}
where $f$ is a smooth function and $w$, $A$ are $1$-forms with
 $\kappa =\frac{1}{2} dA$. (For an elegant proof see \cite{kuenzleduval}.)

For many purposes it is useful to work in {\bf adapted coordinates}: As a
 consequence of the Frobenius theorem for the integrable distribution defined
 by $\tau$, we can introduce local coordinates $(t,x^{1},x^{2},x^{3})$ in the
 neighborhood of any spacetime point such that $\tau = dt$ and
 $\tau (\partial_{i}) = 0$. The integral manifolds are then the slices of
 constant $t$ ({\bf absolute time}). Furthermore, the condition
 $h(\cdot,\tau) = 0$ implies that
 $h = h^{ij} \partial_{i} \otimes \partial_{j}$. In adapted coordinates
 $\lbrace x^{\mu} \rbrace$ with $(x^{0} \equiv t)$
\begin{equation} \label{9}
\tau = dx^{0}, \qquad \qquad  h = h^{ij} \partial_{i} \otimes \partial_{j}
\end{equation}
and the timelike unit vector field 
\begin{equation} \label{10}
V = \partial_{0},
\end{equation}
the expressions for the Christoffel symbols of a symmetric Galilei connection
 become 

\begin{equation} \label{11}
\begin{array}{ll}
 \Gamma^{0}_{\alpha \beta} = 0,  &  \Gamma^{a}_{0b} = h^{ac}(\kappa_{bc} +
 \frac{1}{2} h_{bc,0}) \\
 & \\
 \Gamma^{a}_{00} = 2h^{ab} \kappa_{0b}, &  \Gamma^{a}_{bc}= \frac{1}{2}
 h^{ad}(h_{db,c} + h_{dc,b} - h_{bc,d}). 
\end{array}
\end{equation}
Here $(h_{ij})$ is the inverse matrix of $(h^{ij})$, in other words
 $h^{\flat} = h_{ij} dx^{i} \otimes dx^{j}$. The last equation in (\ref{11})
 proves our previous statement, that the induced connection on the slices of
 constant time is the Levi-Civit\`a connection for the restrictions of
 $h^{\flat}$.

In addition to the {\bf space metric} $h$ we introduce the {\bf time metric}
 $g = \tau \otimes \tau$. Clearly,
\begin{equation} \label{12}
g_{\alpha \beta} h^{\beta \gamma} =0.
\end{equation} 
In contrast to GR, the two (degenerate) metrics $h$ and $g$ are not the
 inverses of each other.

\section{The Newton-Cartan theory}

After these geometrical preparations we can now formulate the Newton-Cartan
 theory in a form which emphasizes its close analogy with GR. The theory
 consists of three parts:

\begin{enumerate}[{\bf I}]
\item Spacetime is a Galilei manifold $(M,h,\tau)$, with a Newtonian
 connection $\Gamma$.
\item Matter is described in part by a symmetric contravariant energy momentum
 tensor $T^{\alpha \beta}$ with vanishing covariant divergence (relative to
 $\Gamma$):
\begin{equation} \label{13}
\nabla _{\beta} T^{\alpha \beta} = 0.
\end{equation}
\item The field equations are 
\begin{equation} \label{14}
R_{\alpha \beta} = 8\pi G(T_{\alpha \beta} - \frac{1}{2} g_{\alpha \beta}T)
 - \Lambda g_{\alpha \beta},
\end{equation}
where $T_{\alpha \beta} := g_{\alpha \sigma} g_{\beta \rho} T^{\sigma \rho}$,
 $T: = g_{\sigma \rho} T^{\sigma \rho}$.
\end{enumerate}
In this formulation we have basically replaced the Lorentz group by the
 Galilei group. Several remarks are in order.

First, it has to be emphasized, that (\ref{13}) is {\em not} a consequence of
 the field equations. This is related to the fact that a Galilei metric does
 not fix the connection.

The specialization to a Newtonian connection lookes somewhat mysterious. There
 is an equivalent formulation of this in terms of a symmetry of the Riemann
 tensor \cite{ehlers3}:
\begin{equation} \label{15}
\alpha (R(\beta^{\sharp}, X)Y) = \beta (R(\alpha^{\sharp}, Y)X)
\end{equation}
for any covectors $\alpha ,\beta$ and vectors $X,Y$; $\sharp$ denotes the map
 $\alpha \mapsto \alpha^{\sharp} = h(\cdot,\alpha)$. In index notation
 (\ref{15}) reads
\begin{equation} \label{15'}
h^{\gamma \rho} R^{\alpha}_{\beta \rho \delta} = h^{\alpha \rho}
 R^{\gamma}_{\delta \rho \beta}.
\end{equation}
In GR, where $h^{\alpha \beta}$ is the inverse of $g_{\alpha \beta}$, this
 symmetry is automatically satisfied. Since the Galilei metric does not fix
 the connection, we have the freedom to impose (\ref{15'}) as a further
 restriction.

The field equations, which can also be written in the form $(g_{\alpha \beta}
 = \tau_{\alpha} \tau_{\beta})$
\begin{equation} \label{16}
R_{\alpha \beta} = 4\pi G \rho \tau_{\alpha} \tau_{\beta} -
 \Lambda \tau_{\alpha} \tau_{\beta}, \qquad \qquad \rho :=T =
 \tau_{\alpha} \tau_{\beta} T^{\alpha \beta},
\end{equation}
allow us to introduce {\bf Galilei coordinates}: Clearly (\ref{16}) implies
 $R^{\alpha \beta} = 0$ and thus by Proposition \ref{prop3} the Galilei
 manifold is locally flat. We can therefore specialize the adapted coordinate
 conditions (\ref{9}) even further such that 
\begin{equation} \label{17}
\tau = dx^{0}, \qquad \qquad h = \delta^{ij} \partial_{i} \otimes \partial_{j}.
\end{equation}
In adapted coordinates we have $R_{ij} = 0 $ as a consequence of the field
 equations, which also implies that the threedimensional time slices are
 locally flat. 

In Galilei coordinates the Christoffel symbols (\ref{11}) simplify to 
\begin{equation} \label{18}
\Gamma^{0}_{\alpha \beta} = 0, \qquad \Gamma^{a}_{00} = 2h^{ac}\kappa_{0c},
 \qquad \Gamma^{a}_{0b}=  h^{ac}\kappa_{bc}, \qquad \Gamma^{a}_{bc} =0.
\end{equation}
The Newton-Cartan theory is slightly more general than Newtons theory of
 gravitation. This can be seen by writing the field equations (\ref{16}) for
 $\Lambda = 0$ in Galilei coordinates. Inserting (\ref{18}) one finds
\begin{equation} \label{19}
2\kappa_{0 \, ,j}^{\ j} - \kappa_{ij}\kappa^{ij} = 4\pi G\rho,
\end{equation}
and
\begin{equation} \label{20}
\kappa_{_i\, ,j}^{\ j} = 0 .
\end{equation}
In addition to this we also have $d\kappa = 0$. We would obtain Newton's 
 theory if the Galilei coordinates could be choosen such that
 $\kappa_{ij} = 0$. (Note that we can still perform time dependent rotations
 and translations.) Now, one can show \cite{ehlers2} that this is possible if
 and only if the following nonlinear condition for the Riemann tensor is
 imposed
\begin{equation} \label{21}
h^{\gamma \rho} R^{\alpha}_{\beta \gamma \delta}
 R^{\beta}_{\alpha \rho \lambda} = 0.
\end{equation}
Relative to Galilei coordinates which satisfy also $\kappa_{ij} = 0$, we
 obtain for $\vec g = -2(\kappa_{01}, \kappa_{02}, \kappa_{03})$ from
 (\ref{19}) and $d\kappa = 0$ the basic equations of the Newtonian theory:
\begin{equation} \label{22}
\mbox{div} \vec g = -4\pi G\rho , \qquad \qquad \mbox{curl} \vec g = 0.
\end{equation}
Ehlers has shown \cite{ehlers2}, that the strange condition (\ref{21}) can be
 deduced from a spatial boundary condition at infinity which can naturally be
 imposed for the description of isolated systems. 

One advantage of the geometrical formulation of the Newton-Cartan theory is
 that the spatial sections can also be chosen as flat tori. This enables us to
 describe spatially compact cosmological models. Some cosmological aspects
 will be presented later.

Finally note that equation (\ref{19}) reads (including the cosmological term)
\begin{equation} \label{23}
\mbox{div} \vec g = -4\pi G \rho + \Lambda + \kappa_{ij}\kappa^{ij}.
\end{equation}
This shows that $\kappa_{ij}\kappa^{ij}$ acts (like a positive $\Lambda$) as a
 repulsive source.

\section{Fluid models in the Newton-Cartan theory}

This section serves mainly as a preparation for our later discussion of 
 Newtonian cosmology.

We introduce again a distinguished timelike unit vector field $V$ on the
 Galilei manifold $(M,h,\tau)$ with time metric $g = \tau \otimes \tau$. The
 integral curves of $V$ define a family of fundamental observers. Note that
 $\tau (V) = 1 $ translates into $\tau_{\alpha} = g_{\alpha \beta} V^{\beta}$.
 The matter model is assumed to be an ideal fluid with four velocity $u$,
 which is also a timelike unit vector field. We begin with some kinematical
 considerations which are familiar in GR.

It is useful to introduce the projection operator
 $P:T_{x}M \rightarrow S_{x}:=\mbox{ker}\tau_{x}$ from the tangent spaces
 onto the horizontal (i.e., spacelike) subspaces definied by 
\begin{equation} \label{24}
P(X) = X - g(X,V)V.
\end{equation}
Clearly, $P(V) = 0$ and $\tau \left( P(X) \right) = 0$. The components of $P$
 are
\begin{equation} \label{25}
P^{\nu}_{\mu} = \delta^{\nu}_{\mu} - g_{\mu \lambda} V^{\lambda}V^{\nu}.
\end{equation}
As before, $h_{\mu \nu}$ denotes the components of $h^{\flat}$. We have the
 identities
\begin{equation} \label{26}
\begin{array}{lll}
P^{\nu}_{\mu} = h_{\mu \lambda}h^{\lambda \nu},  & 
 P^{\mu}_{\lambda} h^{\lambda \nu} = h^{\mu \nu}, &
  P^{\lambda}_{\mu} h_{\lambda \nu} = h_{\mu \nu}, 
 \\ \qquad & \qquad & \\
P^{\lambda}_{\mu}P^{\nu}_{\lambda} = P^{\nu}_{\mu},  & 
 P^{\sigma}_{\rho}P^{\rho}_{\sigma} = 3. & \qquad 
\end{array}
\end{equation}
For the covariant derivatives of $u$ and $V$ one verifies readily the
 following facts:
\begin{equation} \label{27}
\begin{array}{ll}
g_{\mu \lambda} V^{\lambda}_{;\nu} = 0,  &
 \ g_{\mu \lambda}u^{\lambda}_{;\nu} = 0, \\
 \qquad & \qquad \\
\nabla_{X}V \mbox{~and~~} \nabla_{X}u \mbox{~are}  & 
 \mbox{horizontal}, \\
\qquad & \qquad \\
 P^{\mu}_{\lambda} V^{\lambda}_{;\nu} = V^{\mu}_{;\nu},  &
 \ P^{\mu}_{\lambda} u^{\lambda}_{;\nu} = u^{\mu}_{;\nu}.
\end{array}
\end{equation}
The {\bf vorticity} (relative to $V$) is the skew symmetric bilinear form
\begin{equation} \label{28}
\Omega(X,Y) = \frac{1}{2} [h(\nabla_{P(Y)} u, P(X)) - h(\nabla_{P(X)} u, P(Y))]
\end{equation}
and the (rate of) {\bf strain} is 
\begin{equation} \label{29}
\Theta(X,Y) = \frac{1}{2} [h(\nabla_{P(Y)} u, P(X)) +
 h(\nabla_{P(X)} u, P(Y))].
\end{equation}
The {\bf expansion} rate is 
\begin{equation} \label{30}
\theta = h^{\alpha \beta} \Theta_{\alpha \beta}
\end{equation}
and the (rate of) {\bf shear} is the trace-free part of the strain
\begin{equation} 
\sigma(X,Y) = \Theta(X,Y) - \frac{1}{3} \theta h(X,Y).
\end{equation}
While these quantities have the usual interpretation for the fluid motion
 relative to $V$, they are, unfortunatly, {\em not tensoriel}. They are,
 however, simply related to the contravariant tensor fields
 $\Omega^{\sharp}, \Theta^{\sharp}$ with components
\begin{equation} \label{32}
\Omega^{\alpha \beta} = \frac{1}{2} (u^{\alpha}_{;\lambda} h^{\lambda \beta}
 - u^{\beta}_{;\lambda}h^{\lambda \alpha}),
\end{equation}
\begin{equation} \label{33}
\Theta^{\alpha \beta} = \frac{1}{2} (u^{\alpha}_{;\lambda} h^{\lambda \beta}
 + u^{\beta}_{;\lambda}h^{\lambda \alpha}).
\end{equation}
Indeed, the components of (\ref{28}) and (\ref{29}) are given by
\begin{equation} \label{34}
\Omega_{\alpha \beta} = h_{\alpha \rho} h_{\beta \sigma} \Omega^{\rho \sigma},
 \qquad \qquad \Theta_{\alpha \beta} = h_{\alpha \rho}
 h_{\beta \sigma} \Theta^{\rho \sigma}.
\end{equation}
With (\ref{26}) and (\ref{27}) one finds that $\theta$ is simply given by 
\begin{equation} \label{35}
\theta = u^{\sigma}_{;\sigma}
\end{equation}
and the covariant derivative of the velocity field can be decomposed as follows
\begin{equation} \label{36}
h_{\alpha \lambda} u^{\lambda}_{;\beta} = \Theta_{\alpha \beta} +
 \Omega_{\alpha \beta} + h_{\alpha \rho} V^{\lambda} u^{\rho}_{;\lambda}
 g_{\beta \sigma}V^{\sigma}
\end{equation}
With the help of (\ref{36}) we can now derive a Raychaudhuri equation in the
 Newton-Cartan theory. As in GR we start from the identity
\begin{displaymath}
u^{\alpha}_{;\beta ;\gamma} - u^{\alpha}_{;\gamma ;\beta} =
 R^{\alpha}_{\sigma \gamma \beta} u^{\sigma}
\end{displaymath}
which gives
\begin{equation} \label{37}
u^{\beta}u^{\alpha}_{;\alpha ;\beta} = (u^{\beta}
 u^{\alpha}_{;\beta})_{;\beta} - u^{\beta}_{;\alpha}u^{\alpha}_{;\beta}
 - R_{\alpha \beta} u^{\alpha} u^{\beta}.
\end{equation}
With the help of (\ref{36}) and the identities collected in (\ref{26}) and
 (\ref{27}) one can write the second term on the right as follows
\begin{equation} \label{38}
u^{\alpha}_{;\beta}u^{\beta}_{;\alpha} = h^{\rho \alpha} h^{\sigma \beta}
 (\Theta_{\rho \beta} \Theta_{\sigma \alpha} +
 \Omega_{\rho \beta} \Omega_{\sigma \alpha}).
\end{equation}
The first term on the right in (\ref{37}) is
\begin{equation} \label{39}
(u^{\beta} u^{\alpha}_{;\beta})_{;\alpha} = \mbox{div}(\nabla_{u}u).
\end{equation}
After a few steps (see \cite{rueede}), we arrive at the following two
 equivalent forms of the {\bf Raychaudhuri equation}
\begin{equation} \label{40}
\begin{split}
\mbox{div} (\nabla_{u}u)  &  = \nabla_{u}\theta + \frac{1}{3} \theta^{2} +
 h^{\alpha \rho}h^{\beta \sigma} (\sigma_{\rho \sigma} \sigma_{\alpha \beta}
 - \Omega_{\rho \sigma} \Omega_{\alpha \beta}) + \mbox{Ric}(u,u) \\
 &  = \nabla_{u}u + \frac{1}{3} \theta^{2} + h_{\alpha \rho}h_{\beta \sigma}
 (\sigma^{\rho \sigma} \sigma^{\alpha \beta} - \Omega^{\rho \sigma}
 \Omega^{\alpha \beta}) + \mbox{Ric}(u,u).
\end{split}
\end{equation}
Note that these equations hold for any Galilei manifold with a symmetric
 Galilei connection.

At this point we use the field equations (\ref{16}) and obtain (with
 $\tau_{\alpha}\tau_{\beta} = g_{\alpha \sigma} g_{\beta \rho}
 u^{\sigma}u^{\rho}$) 
\begin{equation} \label{41}
\mbox{div} (\nabla_{u}u)   = \nabla_{u}\theta + \frac{1}{3} \theta^{2}
 + h^{\alpha \rho}h^{\beta \sigma} (\sigma_{\rho \sigma} \sigma_{\alpha \beta}
 - \Omega_{\rho \sigma} \Omega_{\alpha \beta}) + 4\pi G\rho - \Lambda.
\end{equation}
This equation will play an important role.

Now we consider an ideal fluid with the energy momentum tensor
\begin{equation} \label{42}
T = \rho u \otimes u + ph.
\end{equation}
From $\nabla T = 0$ one obtains the continuity equation
\begin{equation} \label{43}
\mbox{div} (\rho u) =0
\end{equation}
and the Euler equation
\begin{equation} \label{44}
\nabla_{u}u = -\frac{1}{\rho} \mbox{div} (ph).
\end{equation}
In contrast to GR, equation (\ref{43}) is a conservation law, because it is
 for a symmetric Galilei connection equivalent to ($\mbox{L}_{u}$ denotes the
 Lie derivative with respect to $u$)
\begin{equation} \label{43'}
\mbox{L}_{u} (\rho \mbox{vol}) = 0,
\end{equation}
where vol is the standard volume $\tau \wedge \mbox{vol}_{3}$,
 $\mbox{vol}_{3}$ being the Riemannian volume form of the spatial slices.
 (This equivalence can easily be verified in adapted coordinates.) Thus the
 integral of $\rho \mbox{vol}$ over a comoving domain remains constant.

We mention that it is possible to derive the Raychaudhuri equation (\ref{41})
 also from the Euler equation and the field equation \cite{rueede}. (This is
 closer to what one does in nonrelativistic fluid dynamics.) The two quite
 different derivations reflect some kind of consistency between field and
 matter equations.

As an application of (\ref{41}) we now show, that there are {\em no static
 dust solutions} in the Newton-Cartan theory for $\Lambda =0$ and that for
 $\Lambda >0$ there is just one static solution, which corresponds to the
 {\bf Einstein universe}.

By definition a {\bf static} velocity field $u$ is one with vanishing
 vorticity,
\begin{equation} \label{45}
\Omega^{\sharp} \,
 (=\Omega^{\alpha \beta} \partial_{\alpha} \otimes \partial_{\beta}) =0,
\end{equation}
and for which the Lie derivatives of the expansion and the strain vanish:
\begin{equation} \label{46}
\mbox{L}_{u}\theta = 0, \qquad \mbox{L}_{u}\Theta^{\sharp} = 0 \qquad
 \ (\Theta^{\sharp} = \Theta^{\alpha \beta} \partial_{\alpha}
 \otimes \partial_{\beta}).
\end{equation}

Indeed, assume that there is no pressure term in (\ref{42}), then (\ref{44})
 reduces to $\nabla_{u}u = 0$. Using also the staticity conditions in the
 Raychaudhuri equation (\ref{41}), we find
\begin{equation} \label{47}
4\pi G\rho = \Lambda - h^{\alpha \rho} h^{\beta \sigma}
 \sigma_{\rho \sigma} \sigma_{\alpha \beta} - \frac{1}{3} \theta^{2}.
\end{equation}
This equation has for $\Lambda = 0$ obviously no solution with $\rho >0$.
 (Note, we have not used the second equation of (\ref{46}) to arrive at this
 conclusion.)

Consider next the case $\Lambda >0$. If we write (\ref{47}) in terms of
 Galilei coordinates, we obtain
\begin{equation} \label{48}
4\pi G \rho = \Lambda - \frac{1}{2}[u^{i}_{,j}u^{j}_{,i} +
 \sum_{i,j} (u^{i}_{,j})^{2}].
\end{equation}
In such coordinates one has (with the first equation in (\ref{46})) 
\begin{displaymath}
h^{ij}(\mbox{L}_{u}\Theta^{\sharp})_{ij} = u^{i}_{,j} u^{j}_{,i} +
 \sum_{i,j} (u^{i}_{,j})^{2},
\end{displaymath}
and this vanishes by the second equation of (\ref{46}). Thus the density
 $\rho$ satisfies the relation
\begin{equation} \label{49}
\rho = \frac{\Lambda}{4\pi G}
\end{equation}
of the Einstein universe.

These conclusions hold in particular for Newtonian cosmological dust models.
 It has to be emphasized that we have not made any symmetry assumptions (apart
 from staticity). A very similar argument works also in GR \cite{rueede}.

\section{Newton-Cartan cosmology}

It is very fortunate that the post-recombination universe can be
described
 largely in the Newtonian appro\-ximation. This brings enormous
 simplifications in treating the problems of structure formation, in
 particular in the nonlinear regime. Thanks to this circumstance, we
can for
 instance use N-body simulations.

We consider this as a motivation  (beside others) to put Newtonian
cosmology
 on a conceptually firm basis. This can readily be achieved in the
framework
 of the geometrical formulation of the Newton-Cartan theory that we
have
 described in the previous sections. Again, the analogy to GR is
striking. To
 illustrate this, we consider first homogeneous cosmological models
and then
 develop the cosmological perturbation theory of Friedmann-Lema\^itre models.

\subsection{Homogeneous cosmological models}

In analogy to the discussion of homogeneous cosmological models in GR
(for an
 introduction see \cite{straumann}) we consider first the geometrical
aspect,
 without imposing the field equations. Spacetime is then described by
a
 Galilei manifold $(M,h,\tau)$ with a symmetric Galilei connection
$\Gamma$.
 We introduce adapted coordinates (see equations (\ref{9})). The
spatial
 coordinates $\lbrace x^{i}\rbrace $ parametrize the slices $\sum_{t}$
of
 constant time on which $h$ induces the Riemannian metric $h^{\flat}
= h_{ij}
 dx^{i}  \otimes dx^{j}$. We choose again $V = \partial_{t}$.

Let us assume now that there is a free isometric left action of a
 $3$-dimensional Lie group $G$ on the slices $\sum_{t}$ with $G$ on
which
 $h^{\flat}$ defines a time-dependent family of Riemannian
metrics. Relative
 to a left invariant basis $\lbrace \theta^{a}\rbrace$ of $G$ this
family is
 of the form $h^{\flat} = h_{ab}(t) \theta^{a} \otimes \theta^{b}$.

Using $\nabla g = \nabla h =0$ $(g = \tau \otimes \tau = dt \otimes
dt)$,
 Cartans structure equations for the connection and the Maurer-Cartan
 equations for the Lie group $G$, one can then work out the Ricci
tensor for
 all Bianchi types, with the result given in \cite{rueede}. Here, we
consider
 only the Bianchi type~I, because the field equations imply that the
 $\sum_{t}$ are flat. The metric homogeneity is thus not an additional
 restriction in the Newton-Cartan theory.

For the choice $\theta^{a}= dx^{a}$ we can compute the Ricci tensor
also
 directly with the help of (\ref{11}) and set up the field equations
 (\ref{16}). The result is
\begin{eqnarray}
R_{00} & = & -\frac{1}{2}(h^{ij}\dot h_{ij})_{,0} -
 \frac{1}{4} h^{ij}\dot h_{jk} h^{kl}\dot h_{li}  +
2h^{ij}\kappa_{0j,i} +
 \kappa^{ij}\kappa_{ij}  \nonumber \\
  & = & 4\pi G \rho - \Lambda , \label{50} \\
R_{0i} & = & h^{jk}\kappa_{ik,j} \ =\ 0.   \label{51}
\end{eqnarray}
$R_{ij}$ vanishes identically. Equation (\ref{50}) is obviously
equivalent to
 the Raychaudhuri equation (\ref{41}) for $u=V$. The latter reads in
adapted
 coordinates for any velocity field $u$
\begin{equation} \label{52}
\mbox{div}\nabla_{u}u = \nabla_{u} \theta + \frac{1}{3} \theta^{2} +
 \sigma^{ab}\sigma_{ab} - \Omega^{ab} \Omega_{ab} + 4\pi G \rho - \Lambda .
\end{equation} 
The other field equation (\ref{51}) is equivalent to $\Omega^{ij}\! \!
_{,j}
 =0$ for $u=V$, since for any $u$
 \begin{equation} \label{53}
\Omega_{ab} = \frac{1}{2} [h_{ac}u^{c}_{,b} -h_{bc}u^{c}_{,a} -2\kappa_{ab}].
\end{equation}
We give also the expressions for the other kinematical quantities:
\begin{eqnarray}
\Theta_{ab} & = &  \frac{1}{2}[h_{ac}u^{c}_{,b} + h_{bc}u^{c}_{,a} +
 \dot h_{ab}], \label{54} \\
\theta & = & u^{a}_{,a} + \frac{1}{2} h^{ab}\dot h_{ab} , \label{55} \\
\mbox{div} (\nabla_{u}u) & = & \dot u^{a}_{,a} + 2h^{ab}\kappa_{0a,b}
+
 2h^{bc} \kappa_{ac} u^{a}_{,b} +  h^{bc}u^{a}_{,b} \dot h_{ac} +
 u^{a}u^{b}_{,ab} + u^{a}_{,b}u^{b}_{,a} . \label{56}
\end{eqnarray}
Beside this the $2$-form $\kappa$ is assumed to be closed (Newtonian
 connection).

Matter is assumed to be an ideal fluid with energy momentum tensor
(\ref{42}).
 In adapted coordinates the continuity equation (\ref{43}) and the
Euler
 equation (\ref{44}) become 
\begin{equation} \label{57}
\dot \rho + (\rho u^{i})_{,i} + \frac{1}{2} h^{ij} \dot h_{ij} \rho = 0
\end{equation}
and 
\begin{equation} \label{58}
\dot u^{i} + u^{j}  u^{i}_{,j} + 2h^{ij}\kappa_{0j} + 2u^{j}
(\frac{1}{2}
 h^{ik}\dot h_{kj} + h^{ik} \kappa_{jk} ) + \frac{1}{\rho} h^{ij}p_{,j} = 0 .
\end{equation}
Unlike as in GR, we cannot conclude from our basic equations that the
physical
 quantities like $\rho$ and $p$ are only functions of time. The reason
is
 clear: As already emphasized, the field equations imply that
spacetime has to
 be of Bianchi type~I.
\\\\
Let us specialize the field and matter equations to Newtonian gravity,
 characterized by condition (\ref{21}). We can then introduce Galilei
 coordinates such that $\kappa_{ij} = 0$. Relative to these equations
 (\ref{57}), (\ref{58}) and (\ref{50}) reduce  to
 
\begin{equation} \label{59}
\begin{array}{l}
\dot \rho  + (\rho u^{i})_{,i} = 0,  \\ \\
\dot u^{i} + u^{j}u^{i}_{,j} = -\frac{1}{\rho} p_{,i} + g^{i}, \\  \\
-g^{i}_{,i} = 4\pi G \rho - \Lambda ,
\end{array}
\end{equation}
where  $g^{i}  = -2\kappa_{oi}$ (as in (\ref{22})). We thus arrive at
the
 traditional equations for Newtonian gravity, coupled to an ideal fluid.
\\\\
Let us go back to the Newton-Cartan theory and assume now that $\rho$
and $p$
 are -- in adapted coordinates  -- only functions of $t$. The Euler
equation
 (\ref{44}) implies then $\nabla_{u}u = 0 $ and the continuity
equation
 (\ref{57}) shows that $u^{i}_{,i}$ depends only on $t$. This leads to
the
 following simplification of (\ref{52})
\begin{equation} \label{60}
\dot \theta + \frac{1}{3} \theta^{2} + \sigma^{ab}\sigma_{ab}  -
 \Omega^{ab} \Omega_{ab} + 4\pi G \rho - \Lambda =0.
\end{equation}
Here we have used that $\theta$ is also only a function of $t$,
because
 (\ref{57}) and (\ref{55})  imply 
\begin{equation} \label{61}
\dot \rho + \theta \rho =0.
\end{equation}
Specializing again to Newtonian gravity, we can reach stronger conclusions.

\begin{lemma}
If $\Omega^{\sharp} + \Theta^{\sharp}$  is also translational
invariant, then
 there exist for Newtonian gravity Galilei  coordinates relative to
which the
 spatial components of $u = \partial_{0} + u^{i}\partial_{i}$ are
linear
 functions of the $x^{j}$:
\begin{equation} \label{62}
u^{i} = a^{i}_{j}(t) x^{j} + b^{i}(t).
\end{equation}
\end{lemma}

\begin{description}
\item[Proof:] We know that we can introduce Galilei coordinates such
that $\kappa_{ij} =0$. Since $\mbox{L}_{\partial_{k}}(\Omega^{\sharp}
+ \Theta^{\sharp}) =0$ implies that $(\Omega_{ij} + \Theta_{ij})_{,k}
=0$, equations (\ref{54}) and  (\ref{55}) show that
$u^{i}_{,jk}=0$. This proofs the assertion. 
\end{description}
With a time dependent translation we can pass to Galilei coordinate
for which
 the inhomogeneity in (\ref{62}) disappears. We are now in a situation
which
 has been discussed in classic papers by Heckmann and Sch\"ucking
 \cite{heckmann, heckmannschuecking}.
\\\\
We consider finally homogeneous and isotropic Friedmann-Lema\^itre
models in
 the Newton-Cartan theory. We find these with the ansatz
\begin{equation} \label{63}
V=u, \qquad \qquad  h_{ij} = a^{2}(t) \delta_{ij}, \qquad \qquad 
 \Omega^{\sharp} =0.
\end{equation}
From the remark  connected to equation (\ref{6}) it is clear, that the
 symmetric Galilei connection is now fixed and given by
 $^{V}\Gamma^{\mu}_{\alpha \beta}$. Furthermore, we find
\begin{equation} \label{64}
\theta =  3\frac{\dot a}{a}, \qquad \qquad \sigma_{ab} =0, \qquad
\qquad
 \dot \theta = 3\frac{\ddot a a - \dot a^{2}}{a^{2}},
\end{equation}
and the basic equations (\ref{60}), (\ref{61}) reduce to
\begin{eqnarray}
0 & = & \dot \rho + 3 \frac{\dot a}{a} \rho , \label{65} \\
\ddot a & = & -\frac{4\pi G}{3} \rho + \frac{\Lambda}{3}. \label{66}
\end{eqnarray}
Equation (\ref{65}) implies
\begin{equation} \label{67}
\rho = \rho_{0} \left( \frac{a_{0}}{a} \right) ^{3} ,
\end{equation}
and when this is used in (\ref{66}) we obtain the Friedmann equation  
\begin{equation} \label{68}
 \dot a^{2}  + k  =  \frac{8\pi G}{3} \rho a^{2} + \frac{\Lambda}{3} a^{2},
\end{equation} 
in which the integration constant $k$  can be choosen to be $k = 0,\pm 1$.

In the next section we discuss the perturbation theory of these
homogeneous
 and isotropic solutions in the framework of the Newton-Cartan theory.

\subsection{Cosmological perturbation analysis in the Newton-Cartan theory}

We consider cosmological models deviating only by a small amount from
a
 Friedmann-Lema\^itre  universe, which is defined to be the
background.
 Correspondingly we split all geometric and matter variables into
their
 background values, indexed by $^{(0)}$, and small deviations
 $\delta  p$, $\delta \rho$, $\delta  \kappa$, etc.

The Galilei metric $(h,\tau)$ is kept fixed. This determines the part
 $^{V}\Gamma$, given in (\ref{5}), of the symmetric Galilei
connection.
 Because this is just the background connection, we have $\kappa^{(0)}
= 0$.
 The perturbation of the connection is entirely described by $\delta
\kappa$.
 We also note that $g(u,u) =1$ $(\tau  (u) =1)$ requires that the four
 velocity field is of the form
\begin{equation} \label{69}
u = \partial_{0} + \delta u^{i} \partial_{i}.
\end{equation}
Inserting all this into the field and matter equations leads to a set
of
 perturbation equations for $\delta \rho$, $\delta  p$, $\delta u^{i}$
and
 $\delta  \kappa$ which are still exact. In writing them down, we drop
the
 variational symbol $\delta$ and use the notation
\begin{equation} \label{70}
\rho = \rho^{(0)} (1+D)
\end{equation}
In \cite{rueede} the following complete set of perturbation equations
is
 derived:
\begin{equation} \label{71}
\dot D + [u^{i}(1+ D)]_{,i} =0,
\end{equation}
\begin{equation} \label{72}
\dot u^{i} + u^{j}u^{i}_{,j} + 2\frac{\dot a}{a}u^{i} =
 -\frac{1}{\rho}h^{ij}p_{,j} - 2h^{ij}\kappa_{0j} + 2h^{ij}\kappa_{jl}u^{l},
\end{equation}
\begin{equation} \label{73}
2h^{ij}\kappa_{0j,i} = -\kappa^{ij} \kappa_{ij} + 4\pi G\rho^{(0)} D,
\end{equation}
\begin{equation} \label{74}
d\kappa = 0,
\end{equation}
\begin{equation} \label{75}
h^{jl}\kappa_{ij,l} = 0.
\end{equation}
These agree for $\kappa_{ij} = 0$ with the usual Newtonian
perturbation
 equations (see, e.g., \cite{peebles}). (In making this comparison one
has to
 note, that the peculiar velocity field $v^{i}$  is usually defined by
 $v^{i} = a(t)u^{i}$. The gravitational field is again given by
$\kappa_{0i}:$
 $g^{i} = -2\kappa_{0i}$.)

Linearization of the perturbation equations gives
\begin{equation} \label{76}
\dot D + u^{i}_{,i} = 0,
\end{equation}
\begin{equation} \label{77}
\dot u^{i} + 2\frac{\dot a}{a} u^{i} =
-\frac{1}{\rho^{(0)}}h^{ij}p_{,j} -
 2h^{ij}\kappa_{0j},
\end{equation}
\begin{equation} \label{78}
2h^{ij}\kappa_{0j,i} = 4\pi G\rho^{(0)}  D,
\end{equation}
\begin{equation} \label{79}
d\kappa = 0, \qquad \qquad h^{jl}\kappa_{ij,l} = 0.
\end{equation}
Eliminating $u^{i}$ we arrive at the well-known  perturbation equation
for the
 density fluctuations:
\begin{equation} \label{88}
\ddot D + 2\frac{\dot a}{a}D = \frac{1}{\rho^{(0)}}h^{ij}p_{,ij} +
 4\pi G \rho^{(0)}D.
\end{equation}
From (\ref{71}) - (\ref{75}) we can derive in a standard manner
(exact)
 perturbation equations for vorticity and shear. One equation  agrees
with the
 Raychaudhuri equation for the perturbations:
\begin{equation} \label{89}
\nabla_{u}\theta + \Theta^{ij}\Theta_{ij} - \Omega^{ij}\Omega_{ij} +
 \left( \frac{1}{\rho}h^{ij}p_{,j} \right) _{,i}  + 4\pi G\rho - \Lambda = 0.
\end{equation}
For the vorticity one finds \cite{rueede}
\begin{equation} \label{90}
(\nabla_{u}\Omega)_{ij} + h^{\alpha  \beta}\Omega_{\beta j}\Theta_{i
\alpha} -
 h^{\alpha \beta}\Omega_{\alpha i} \Theta_{\beta j} = p_{,[j}
 \left( \frac{1}{\rho}\right)\, _{,i]}.
\end{equation}
(We have again dropped the variational symbol on $\Omega$, $\Theta$,
$u$, $p$;
 but $\rho$ is the total density.)

We conclude this discussion by writing the exact perturbation
equations
 (\ref{71}) - (\ref{75}) in a covariant form:
\begin{equation}   \label{91}
\nabla_{V}D + \mbox{div}[(1 + D)(\delta^{\beta}_{\alpha}  -
 g_{\alpha \lambda}V^{\lambda}V^{\beta})u^{\alpha}\partial_{\beta}] = 0,
\end{equation}
\begin{equation} \label{92}
\nabla_{u}u = -\frac{1}{\rho}\mbox{div}ph,
\end{equation}
\begin{equation} \label{93}
\mbox{div}\nabla_{V}V = -h^{\alpha \gamma}h^{\beta \delta} 
 \kappa_{\alpha \beta}\kappa_{\gamma \delta} + 4\pi G\rho^{(0)}D,
\end{equation} 
\begin{equation} \label{94}
d\kappa = 0,
\end{equation}
\begin{equation} \label{95}
h^{\alpha \beta}\nabla_{\alpha}[(\delta^{\sigma}_{\gamma}  - g_{\gamma
\delta}
 V^{\delta}V^{\sigma}) \kappa_{\sigma \beta}] = 0.
\end{equation}

\section{Concluding remarks}

The Newton-Cartan theory can sometimes provide useful insights for
problems in
 GR. An interesting example concerns the cosmic no-hair conjecture,
which is
 not yet settled in sufficient generality within GR. Bauer et al
\cite{bauer} 
 were, however, able to prove satisfactory theorems in the framework
of the
 Newton-Cartan theory. For ideal fluid models they showed that
solutions
 corresponding to nearly homogeneous initial data for a compact time
slice
 exist in the case $\Lambda >0$ for all positive times and that the
difference
 between the inhomogeneous and homogeneous solutions tends to zero in
a strong
 sense. Perturbations are thus strongly damped. Presumably a
corresponding
 nonlinear stability property holds also in GR, but this appears very
 difficult to prove.

The geometrical formulation of the Newton-Cartan theory has also
played a
 useful role in rigorous discussions of the Newtonian limit of GR
 \cite{ehlers2}. The starting point is the observation by Ehlers that
both
 theories fit naturally into a larger frame theory with two metrics
 $h^{\alpha \beta}$, $g_{\alpha \beta}$ related by
 $g_{\alpha \sigma} h^{\sigma \beta} = -\lambda
\delta^{\beta}_{\alpha}$ 
 ($\lambda = 1$ for GR and $\lambda =0$ for the Newton-Cartan  theory).

This frame theory has also played a remarkable role in the work of
Heilig
 \cite{heilig} for establishing rigorous existence theorems in GR for
 solutions which describe rotating stars.

Several more formal aspects have been studied by K\"unzle and
collaborators.
 An example is the genera\-lization of the Galilei invariant
 spin-$\frac{1}{2}$-wave equation to a curved Newton space-time
 \cite{kuenzleduval, kuenzleduval2}.

Finally, without being complete, we mention that Julia and Nicolai
 \cite{julianicolai} have recently obtained the Newton-Cartan theory
through a
 dimensional reduction of a Kaluza-Klein theory along a null vector.

All this demonstrates once more the remarkable continuity in the
development
 of theoretical physics. The word ``revolution'' rarely deserves to be
used in
 this context. To our knowledge, it appears in Einsteins writings only
once,
 namely in connection with his hypothesis of the light quantum
 \cite{einstein2}. He did not regard its use to be appropriate in all
his work
 on special and general relativity.
\\\\
{\bf Acknowledgments}
\\\\
One of us (N.~S.) would like to thank Markus Fierz for very
interesting
 correspondence. We are grateful to Marcus Heusler for a careful
reading of
 the manuscript.

\end{document}